\documentclass[%
reprint,
superscriptaddress,
 amsmath,amssymb,
 aps,
]{revtex4-2}

\usepackage{graphicx}% Include figure files
\usepackage{dcolumn}% Align table columns on decimal point
\usepackage{bm}% bold math

\usepackage[usenames,dvipsnames]{color}
\usepackage{graphicx, microtype}
\usepackage[bookmarks=false,colorlinks]{hyperref}
\hypersetup{linkcolor=Plum,citecolor=RubineRed,filecolor=Plum,urlcolor=PineGreen}
\usepackage{soul} 
\usepackage{tabularx}
\usepackage{float}
\usepackage{graphicx}
\usepackage{caption}
\usepackage{ulem}
\usepackage{xspace}
\begin{document}

\title{Energy Landscape analysis of metal-insulator transitions: theory and application to Ca$_2$RuO$_4$, $R$NiO$_3$ and their heterostructures}
% Force line breaks with \\

\author{Alexandru B. Georgescu}
\affiliation{Northwestern University, Department of Materials Science and Engineering, Evanston, Illinois, 60208}
\affiliation{Center for Computational Quantum Physics, Flatiron Institute, 162 5th Avenute, New York, NY 10010}
\email{alexandru.georgescu@northwestern.edu}
\author{Andrew J. Millis}

\affiliation{Center for Computational Quantum Physics, Flatiron Institute, 162 5th Avenute, New York, NY 10010}
\affiliation{Department of Physics, Columbia University, 538 West 120th Street, New York, New York 10027}%

%\affiliation{% Authors' institution and/or address\\This line break forced with \textbackslash\textbackslash}%

\date{\today}% It is always \today, today,

\begin{abstract}

We present a general methodology that enables the disentanglement of the electronic and lattice contributions to the metal-insulator transition by building an energy landscape from numerical solutions of the equation of state. The methodology works with any electronic structure method that provides  electronic expectation values at given atomic positions.  Applying the theory to rare-earth perovskite nickelates ($R$NiO$_3$) and Ruddlesden-Popper calcium ruthenates (Ca$_2$RuO$_4$) in bulk, heterostructure and epitaxially strained thin film forms using equation of state results  from  density functional plus dynamical mean field calculations we show that the electron-lattice coupling is an essential driver of the transition from the metallic to the insulating state in these materials.

\end{abstract}

\maketitle

%\tableofcontents
\section{Introduction \label{sec:introduction}}

The relative importance of electronic and lattice effects in driving phase transitions in quantum materials is a subject of long-standing interest and controversy. One issue of particular  focus is the metal to insulator transition (MIT), which in many materials is accompanied by a lattice distortion \citep{Imada1998}. Multiple studies have addressed various aspects of the interplay between electronic and lattice  contributions to the metal-insulator transition both theoretically and experimentally \citep{Claribel,Georgescu2019,Oleg,Disa2017, Lauren2019,Han2018,Disa2017,Kyle,Karin2021,Mercy2017,JenniferThesis,SohrabPicoscale,Superlattices2011,Kumah2014a,Middey2018,Gray2011,Stemmer2013,AlexEELS,Kotiuga2019,Stemmer2013,Zhang2016b,Shamblin2018,Meyers2016,Forst2017,Forst2015,Medarde1998,Hu2016,Caviglia2013,guzman,AndyArxiv,Park2016,OpticalAntoine,Kumah2014a,polarizationantoine,Han2011, Seth2017, Subedi2015, OpticalAntoine, Strand2014,Blanca-Romero2011,held,Haule2017Nickelates,SohrabPicoscale,DivineAlexNPJ,AlexDivine20202,JMReview,Claribel, BondAngles,Octahedra2,Georgescu2019,VO2Length,Basov2018,VO2Length,Basov2018,Haule2017Nickelates,Uri1,Uri2}. Machine learning approaches are discovering new features from comparison of MIT and non-MIT materials \citep{Georgescu2020,Raymond2020}. Yet, the relative importance of electronic and lattice effects in producing a transition has remained unclear, in part because of the lack of an appropriate theoretical framework and  computational schema for evaluating the needed quantities. This paper aims to rectify both issues.

\begin{figure}[b]
    \centering
    \includegraphics[width=0.45\textwidth]{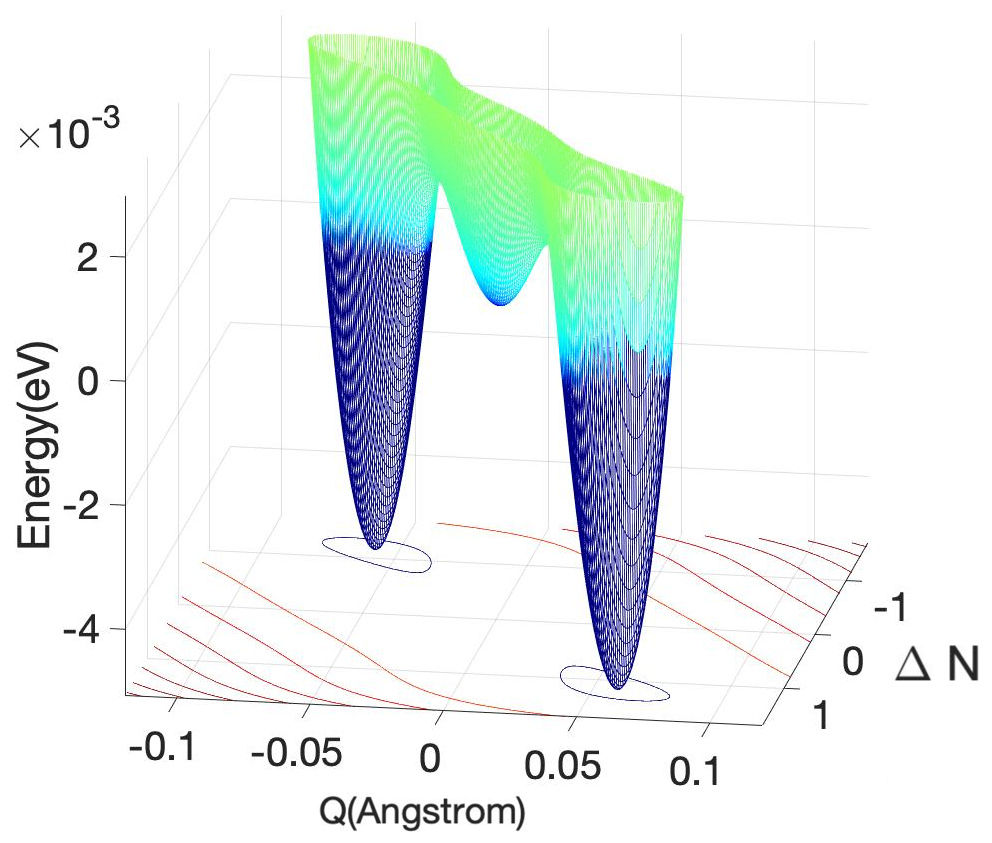}
    \caption{Energy landscape for NdNiO$_3$ as a function of electronic charge disproportionation $\Delta N$ and lattice distortion $Q$ as obtained from the  energy functional methods  introduced in this paper using data from Ref. \citep{Georgescu2019}.}
    \label{fig:rnoE1}
\end{figure}

Metal insulator transition phenomena are properly analysed by constructing an energy landscape in the space of lattice distortions and electronic orders such as that shown in Fig.~\ref{fig:rnoE1}. We show how to build these energy landscapes from numerical data of the kind now available from modern quantum many-body methods and we apply the theory to bulk, thin film and superlattice heterostructures of $R$NiO$_3$ and Ca$_2$RuO$_4$, two material families that exhibit metal-insulator transitions but have different local physics. 

Our work  was inspired by the pioneering work of Fisher and collaborators\citep{FisherEnergy}, who combined elegant experimental measurements with an insightful free energy analysis to show that the observed second order `nematic' transition in an iron pnictide material had an intrinsically electronic origin. Our work may be viewed as a generalization of the ideas of Fisher et. al to first order transitions, where a more global view of the energy landscape is required, and as an extension of the insightful equation of state analysis of Peil, Hampel, Ederer and Georges \citep{Oleg} to the computation and interpretation of the full free energy landscape. We  make substantial use of concepts put forward in  analyses of the metal-insulator transition in epitaxially strained Ca$_2$RuO$_4$  \citep{Han2018} and in  nickelate heterostructures \citep{Georgescu2019}.

The rest of this paper is organized as follows: the Framework section (\ref{sec:framework}) shows how we build total energy landscapes. In sections \ref{sec:ReNiO3} and ~\ref{sec:CaRuO}  we use the formalism to analyse the metal-insulator transitions occurring in two representative classes of materials:  the rare earth nickelate perovskite RNiO$_3$ compounds, where the transition involves a symmetry breaking lattice distortion and two sublattice electronic  ``charge order" and Ca$_2$RuO$_4$ where the transition is isostructural (no structural symmetry breaking) and involves electronic ``orbital order". Section ~\ref{sec:Tdep} presents some qualitative considerations related to the thermally driven transitions. Section ~\ref{sec:summary} is a summary and conclusion, assessing the work and outlining future directions.

\section{Framework \label{sec:framework}}
Following Refs.~\citep{Han2018,Oleg,Georgescu2019} we consider a free energy $F(\Delta N ,Q )$ that depends on an electronic order parameter, $\Delta N$, and a lattice distortion, $Q$. The precise definitions of $\Delta N$ and $Q$ will depend on the specifics of the system. We choose $(Q,\Delta N )=(0, 0)$ as the equilibrium configuration of the metallic phase. The insulating state then corresponds to a configuration $(\Delta N, Q)\neq (0,0)$ and the question of the existence of a purely electronic transition relates to the properties of $F$ as a function of $\Delta N$ at $Q=0$.

A second order transition corresponds to a linear instability of the $(\Delta N,Q)=(0,0)$ state, in other words to a change in sign of the smallest eigenvalue of the Hessian matrix $\partial^2 F/\partial^2 \Delta N,Q$ evaluated at $\Delta N=Q=0$. In general the eigenvector associated with the negative eigenvalue will have components along both $\Delta N$ and $Q$, indicating that the electronic and lattice orders are coupled. A purely electronically driven transition would correspond to a change in sign of $\partial^2F/\partial^2\Delta N$, a lattice-driven transition would correspond to a change in sign of $\partial^2F/\partial^2Q$, and in a mixed situation neither `pure' derivative changes sign but the lowest eigenvalue of the Hessian matrix does change sign. In the case of the nematic transition in pnictides, Fisher and co-workers were able \citep{FisherEnergy} to experimentally determine $\partial^2 F(Q=0)/\partial \Delta N^2$ and show that it changed sign at a temperature only slightly lower than the observed nematic transition temperature, thereby establishing that a purely electronically driven nematic transition exists in these materials, and is simply slightly enhanced by coupling to the lattice.

The transitions of interest in this paper are  first order, characterized by the appearance of a new free energy minimum at which  $\Delta N$ and $Q$ are both different from zero (see e.g. Fig.~\ref{fig:rnoE1}). Study of first order transitions requires global knowledge of the free energy. One may say that the transition is electronically driven if  $F(\Delta N,Q=0)$ has an extremum at a $\Delta N\neq 0$ with $\partial^2 F(\Delta N,0)/\partial \Delta N^2 >0$ and with energy lower than $F(0,0)$.  One may say that the transition is lattice-assisted if $F(\Delta N,0)$ has a  local minimum at a $\Delta N\neq 0$ but that the lattice coupling is required to make this minimum the ground states. Finally if $F(\Delta N,Q=0)$ has only one minimum, at $\Delta N=0$ but exhibits a stable minimum with $\Delta N\neq 0$ at a $Q\neq 0$ the we may say that the transition is not electronically driven. The energy functional shown in Fig. ~\ref{fig:rnoE1}  shows a transition that is not electronically driven  because the global minima are at a $Q,\Delta N\neq 0$ but along the line $Q=0$ the functional has only one minimum, at the origin. It is important to emphasize that this distinction is separate from the question of the importance of correlation physics. It is common in the literature to distinguish strongly from weakly correlated materials according to whether or not current implementations of density functional theory capture the physics; in current language whether a density functional calculation of $F_{el}$ suffices; even if the transition is not electronically driven in the sense given above, beyond density functional correlations might be needed to obtain an $F_{el}(\Delta N,Q)$ with an insulating minimum at the physical $Q$. 

The global knowledge of $F$ required to decide this issue has been challenging to obtain. Available quantum many-body  methods can with reasonable computational cost obtain reasonable estimates of the optimal $\Delta N$ that minimizes $F$ at fixed atomic positions ($Q$), but finding the dependence of $F$ on $\Delta N$ at fixed $Q$ or on $Q$ at fixed $\Delta N$ is difficult, both because it is not straightforward to control $\Delta N$ independently of $Q$ and because calculations of energies are very expensive.

%\ajm{{\bf not sure we need this but lets keep it for now} The continuing development of correlated electron theory including the density functional plus dynamical mean field method  has yielded methodologies for calculating energies \citep{Park2014} and forces \citep{Haule2016ForcesApproach} {\bf vollhardt group had a force ref}. This work has led for example to reasonable estimates for the  dependence of the metal insulator transition in Ruddlesden-Popper ruthenates on epitaxial strain and for the dependence on rare earth ion of the critical pressure for the metal-insulator transition in perovskite nicklates \citep{Park2016}. But because it is theoretically difficult to control $\Delta N$ obtaining the energy landscape has not been possible.}\abg{I think this is important: getting E as a function of $\Delta N$ separately from Q is the main point.}\ajm{Yes but this point is made at the end of the previous paragraph. The question is whether we need to discuss the fact that people have calculated energies and forces}

Here we build on the observation that the metal-insulator transitions of greatest current experimental interest share two features that greatly simplify an analysis. First, the lattice energetics  is to good approximation harmonic \citep{Han2018,Georgescu2019,Oleg}, as shown from calculations \citep{Georgescu2019} and  by the observation that phonon frequencies change only very slightly ($\sim 1\%$) across the metal-insulator transition. Second, and closely related to the first point,  the coupling between the electronic order parameter and lattice modes is linear and is well determined by standard theoretical methods  \citep{Georgescu2019,Han2018,Oleg}, as are the lattice energetics. This means that we may write, schematically, \citep{Oleg,Georgescu2019,Han2018}:

\begin{equation}
    F(\Delta N,Q)=\frac{1}{2}KQ^2-\frac{1}{2}g Q \Delta N +F_{el}(\Delta N)
    \label{Fansatz}
\end{equation}

Requiring stationarity of $F$ with respect to variations in $\Delta N$ and $Q$ yields the equations of state \citep{Oleg,Georgescu2019}
\begin{equation}
 KQ=\frac{1}{2}g\Delta N
\label{Qeq}   
\end{equation}
and 
\begin{equation}
\frac{\partial F_{el}(\Delta N)}{\partial N}=\frac{1}{2}gQ
\label{eqofst}
\end{equation}

Because $g$ and $K$ are known, Eq.~\ref{Qeq} can be solved. Substituting the solution back into Eq. ~\ref{Fansatz} produces: 
\begin{equation}
    \bar{F}(\Delta N)=-\frac{1}{8K}\left(g\Delta N\right)^2 +F_{el}(\Delta N)
    \label{Fansatz1}
\end{equation}
From this point of view the key question is the magnitude of the 'lattice stabilization energy' $\frac{1}{8K}\left(g\Delta N\right)^2$: is it large enough to create a new extremum at a $\Delta N\neq 0$? Is it large enough to ensure that the extremum at $\Delta N=0$ is unstable? Answering this question requires knowledge of $F_{el}$.

Peil, Hampel, Ederer and Georges observed \citep{Oleg} that existing many-body methods such as the dynamical mean field method  enable  computation of electronic properties  at fixed lattice configuration $Q$, in effect solving Eq.~\ref{eqofst}. Here we take one further step:  the  ansatz for the electronic energy functional in Eq. ~\ref{Fansatz} means that the computed $\Delta N(Q)$ determines $dF_{el}(\Delta N)/d\Delta N$; the derivative can then be integrated, thereby constructing $F$ itself. In practice we perform the integration by fitting the $\Delta N(Q)$ results to a polynomial, which is  analytically integrated. While this method requires a choice of polynomial form, we empirically find that the uncertainty thereby introduced is small, of the order of a few meV, comparable to the error in performing an explicit energy calculation. This approach  is  computationally  inexpensive, relatively unaffected by noise, and the analytically specified functional forms provide additional insight, in particular enabling straightforward examination of the global energy landscape. 

We use this approach to analyse the metal-insulator transitions occurring in the rare-earth nickelates and their heterostructures, and in bulk and epitaxially strained Ca$_2$RuO$_4$. We find that in all of these situations the metal-insulator transition is, in the sense defined above, not electronically driven.

\section{Bulk Rare-Earth Nickelates  \label{sec:ReNiO3}}

\subsection{Materials and notation}
\begin{figure}[ht]
    \centering
    \includegraphics[width=0.4\textwidth]{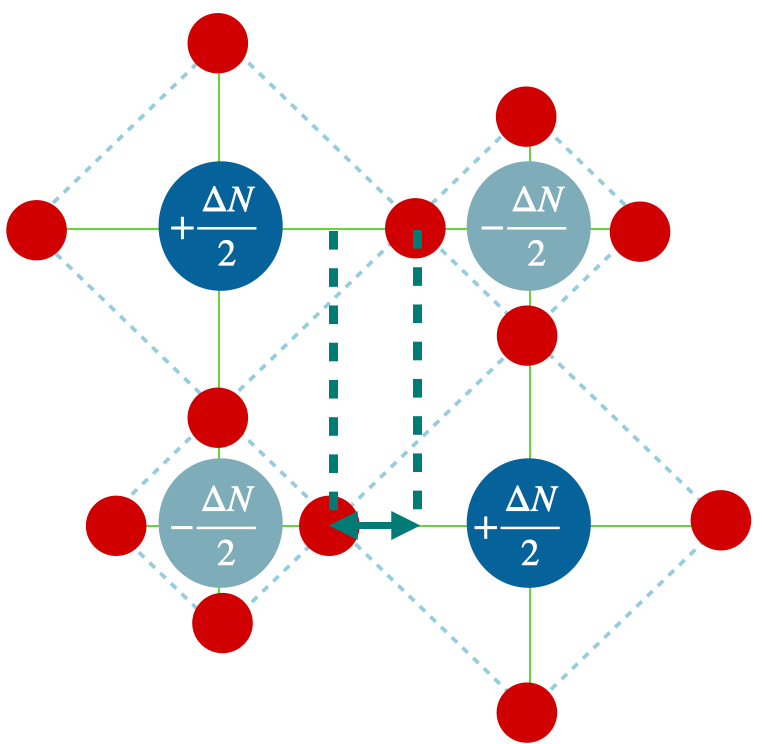}
    \caption{Schematic of the electronic and structural disproportionation in the insulating state in rare-earth nickelates: NiO$_6$ octahedra disproportionate electronically and structurally in a checkerboard pattern. In the low-energy model of Ref.~\citep{Oleg}, the Ni ions corresponding to the larger octahedra have 1e$^-+\frac{\Delta N}{2}$ electrons in their e$_g$ shell, while those corresponding to the smaller one have 1e$^- -\frac{\Delta N}{2}$. The difference in average bond lengths defines $Q$ for the nickelate materials. }
    \label{fig:rno}
\end{figure}

The perovskite rare-earth nickelates have chemical formula $R$NiO$_3$; when synthesized in bulk form  the materials with $R$=Lu, Y, Sm, Nd, Pr exhibit a first order transition from a high temperature metal to a low temperature insulator while LaNiO$_3$ remains metallic down to lowest temperature.  The material properties depend systematically on the choice of rare-earth ion, with  LuNiO$_3$ exhibiting the highest metal insulator transition temperature $T_{MIT}=600 K$, SmNiO$_3$ having $T_{MIT}\approx 400K$, NdNiO$_3$ having a $T_{MIT}\approx 240K$ and PrNiO$_3$ having a $T_{MIT}\approx 130K$\citep{Nick1999,Nick2010,JenniferThesis}. The materials may also be grown in heterostructure form \citep{Disa2017,Georgescu2019,Claribel,Chen2021,Chen2013a,Octahedra2,AlexEELS,SohrabPicoscale,JenniferThesis}, with a small number of layers of $R$NiO$_3$ sandwiched between layers of other materials. The metal-insulator transition temperature in the heterostructures differs from that in bulk. 

Except for the La compound,  where the symmetry is slightly different,  the high temperature metallic state forms in a Pbnm structure  which for present purposes may be regarded as a pseudo-cubic lattice of Ni ions with an O ion at the midpoint of each Ni-Ni bond.  The insulating state is characterized by a two-sublattice bond disproportionation, with one Ni sublattice having a short mean Ni-O bond length and the other by a long one, represented qualitatively in Fig. ~\ref{fig:rno}. The difference in mean Ni-O bond lengths defines the lattice mode $Q$ appropriate to this transition.

In the low T insulating state the two Ni sublattices differ in electronic configuration. This difference has been characterized in various ways in the literature.  \citep{Park2012,Park2013,Park2014,Park2016,Oleg,Oleg,Haule2017Nickelates}. We follow Peil and collaborators \citep{Oleg} and define the electronic change $\Delta N$  as the difference in the occupation of  the e$_g$  states  obtained from a  narrow window Wannier fit of the  bands crossing  the Fermi level.

\subsection{Results, bulk nickelates\label{subsec:bulk}}
\begin{figure}[ht]
    \centering
    \includegraphics[width=0.45\textwidth]{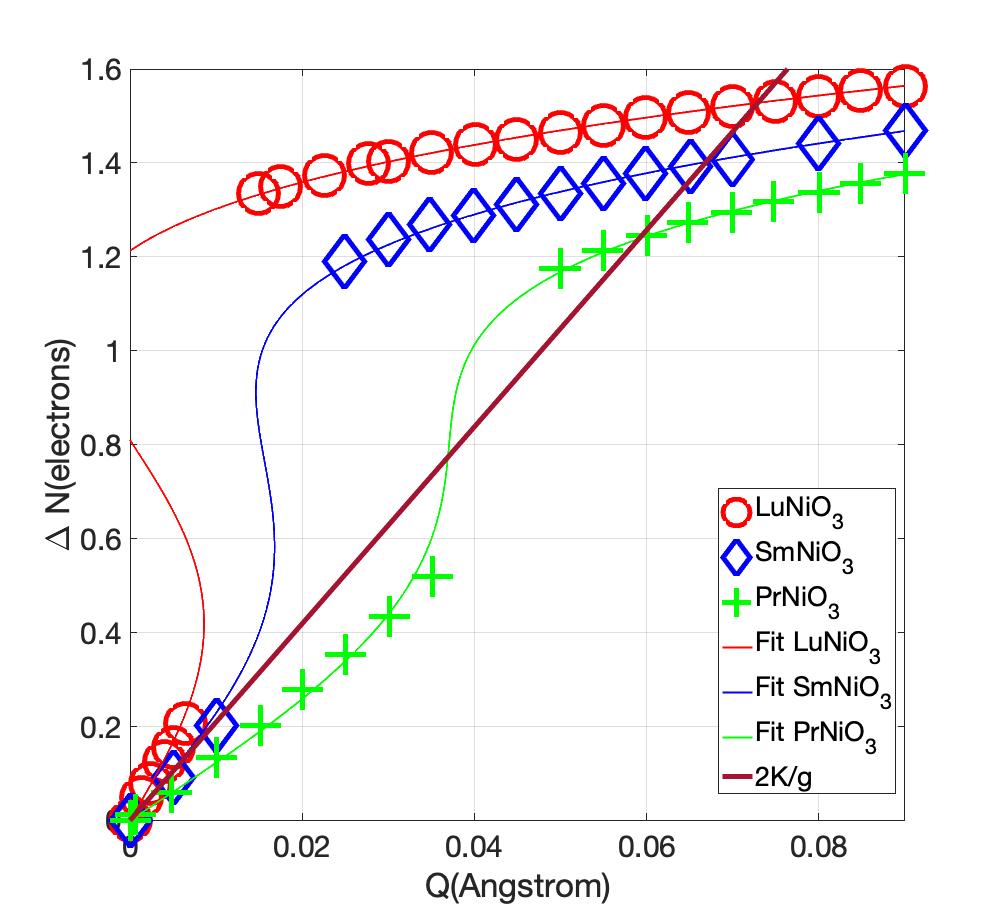}
    \caption{Points: $\Delta N(Q)$  values used for the three $R$NiO$_3$ materials as obtained from DFT+DMFT calculations from Ref.~ \citep{Oleg};  Light lines: fits of the points to  Eq. ~\ref{EOSnickelate}. Also shown is the  $Q(\Delta N)$ line obtained from the $Q$ equation of state Eq.~\ref{Qeq}. Note that  K/g has the same value for all three materials.}
    \label{fig:rno1}
\end{figure}

In Fig. ~\ref{fig:rno1} we present  results  for three different bulk-form rare earth nickelates, (LuNiO$_3$, SmNiO$_3$, PrNiO$_3$) as a function of $Q$, digitized from the calculations of  $\Delta N$   presented in   Ref. ~\citep{Oleg} . In all compounds, two regimes are  found, a low $Q$ regime where $\Delta N$ is proportional to $Q$ and a high $Q$ regime where $\Delta N$ is large but weakly dependent on $Q$; the two regimes are separated by a discontinuity. 

To determine the free energy model to which the results should be fit we observe that the two sublattice nature of the insulating state means that the free energy must be invariant under simultaneous inversion $Q\leftrightarrow -Q$, $\Delta N \leftrightarrow -\Delta N$, so in particular $F_{el}(\Delta N)$ must be even in $\Delta N$. Motivated by the observed first order nature of the transition we assume that $F_{el}$ is a $6^{th}$ order polynomial in $\Delta N$:

\begin{equation}
    F_{el}(\Delta N)=\frac{1}{2}\chi_0^{-1}\Delta N^2+\frac{1}{4}\beta\Delta N^4+\frac{1}{6}\gamma\Delta N^6
    \label{Fel}
\end{equation}

Solving Eq. ~\ref{Qeq} and eliminating $Q$ from the full free energy then changes the quadratic term $\chi_0^{-1} \rightarrow \chi_0^{-1} -\frac{g^2}{4K}$. Note that the shift in the quadratic term involves the parameter $g^2/K$, which varies slightly  across the $R$NiO$_3$ series although as noted in Ref.~\citep{Oleg} the ratio $g/K$ is almost material-independent. The K for the Lu, Sm, Pr materials were reported in reference \citep{Oleg} to be 39.29, 41.45 and 44.47 eV/ \AA$^2$, while the g are 3.75, 4.02 and 4.24 ev/ \AA. 

Eq. ~\ref{Fel} implies that the equation of state Eq. ~\ref{eqofst} takes the explicit form 
\begin{equation}
   0=-gQ+\chi_0^{-1}\Delta N +\beta \Delta N ^3 + \gamma \Delta N^5
   \label{EOSnickelate}
\end{equation}

We fit the points in Fig. ~\ref{fig:rno1} to Eq.~\ref{EOSnickelate}, using the $g$ values presented in Ref.~\citep{Oleg},  obtaining the light solid curves  shown in Fig \ref{fig:rno1} and the fit parameters given  Table \ref{table:rno1}. It is important to note that the DMFT equation of state shown in Fig. ~\ref{fig:rno1} has two branches, with the solution discontinuously changing from one to the other as $Q$ is varied. In our fits  we only use the data points shown as symbols in  Fig.~\ref{fig:rno1}; these points lie outside the region of multistability (i.e. where two $\Delta N$ solutions exist within DMFT for the same Q). However, our theoretical energy function reproduces correctly behavior that was not part of the original fit, for example the locally stable large $\Delta N$ at $Q=0$ solution for  LuNiO$_3$ \citep{Oleg}.   Including points within the region of multistability does not lead to significant changes to our results. The free energy we obtain here may be thought of as the free energy that would be obtained form the two component Landau theory of Ref.~\citep{Oleg} if the metal-insulator variable is integrated out. 

\begin{center}
\begin{table}
\begin{tabular}{ | c | c | c |  c | c | c | c |}
 \hline
 $R$NiO$_3$  & $\chi_0^{-1}$(eV) & $\beta $(eV) & $\gamma $(eV) & $-\frac{1}{4}\frac{g^2}{k}$(eV) \\
 \hline
 LuNiO$_3$ & 0.0596 & -0.1314 & 0.0618 & -0.0895  \\ 
 \hline
 SmNiO$_3$ & 0.0945 & -0.1303 & 0.0667 & -0.0975  \\  
 \hline
 PrNiO$_3$ & 0.1749 & -0.1713 & 0.0804 & -0.1011  \\  
 \hline
\end{tabular}
\caption{Parameters that characterize the electronic and total energy of the three $R$NiO$_3$ materials, as extracted by fitting the data from Ref \citep{Oleg} to Eq.~\ref{EOSnickelate}.}
    \label{table:rno1}
    \end{table}
\end{center}

\begin{figure*}
    \centering
    \includegraphics[width=0.9\textwidth]{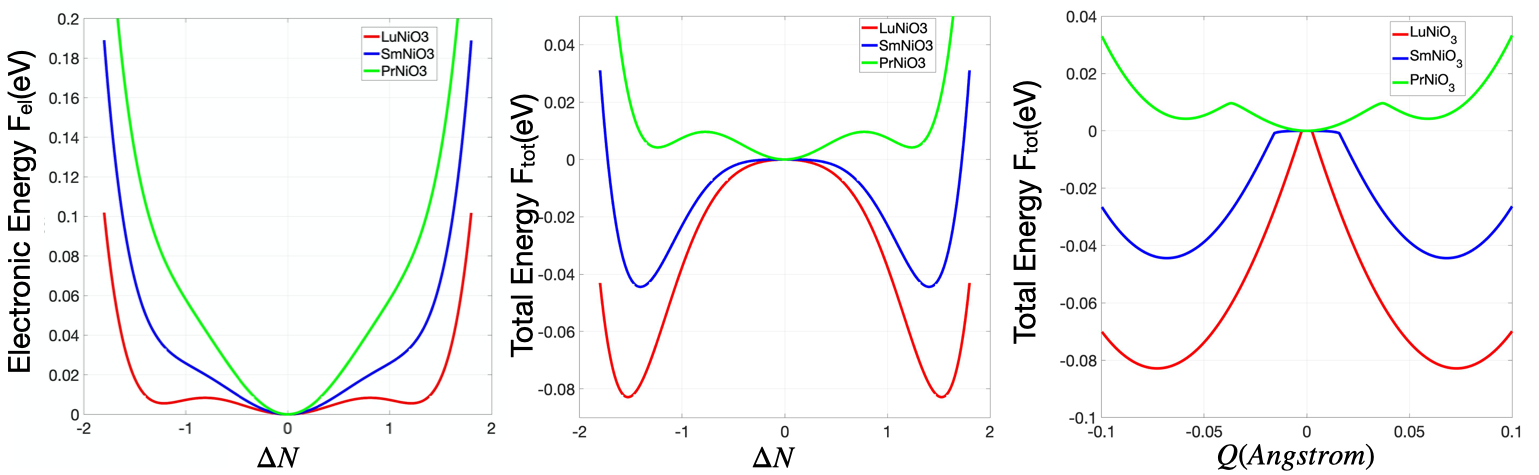}
    \caption{Left panel: electronic energy; central panel: total energy of three RNiO$_3$ compounds as function of electronic order parameter $\Delta N$. Right panel: total energy as function of lattice displacement $Q$.}
    \label{fig:rnoE}
\end{figure*}

The left panel of Fig.~\ref{fig:rnoE} shows the purely electronic energy $F_{el}$ corresponding to the fit parameters given in Table ~\ref{table:rno1}. We see that for all of the materials the purely electronic theory is characterized by a locally and globally stable metallic minimum at $Q=0$. Only for LuNiO$_3$ does a second, insulating extremum exist and even in this case it is not energetically favored.  The central panel presents the total free energy as a function of $\Delta N$ obtained by minimization over the lattice degrees of freedom $Q$. We see that inclusion of the  lattice coupling is necessary to stabilize the $\Delta N,Q\neq0$ solution: in other words, it is the electron-lattice coupling that drives the transition.  

Table ~\ref{table:rno1} shows that as one moves across the RNiO$_3$ series from $R$=Lu (most insulating) to $R$=Pr (least insulating), the linear response electronic susceptibility of the metallic state decreases, consistent with the empirical observation that the Lu material is the strongest insulator and the Pr material is the weakest. We see also that the nonlinear terms in the energy are the same for the Lu and Sm compounds, which are the two strong insulators, but are different in the Pr compound, which is close to the metal-insulator transition. To understand composition-dependent changes one must consider more than just the variation in linear response susceptibilities. Results to  be presented below for superlattices further confirm this point. 

\begin{figure}[b]
    \centering
    \includegraphics[width=0.5\textwidth]{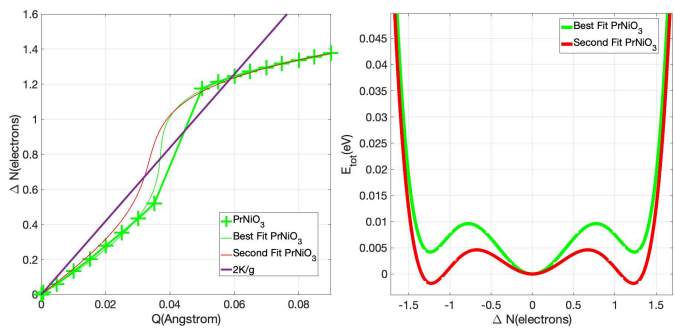}
    \caption{Fits and total energy for PrNiO$_3$ based on DMFT data from Ref \citep{Oleg}; green lines correspond to the optimal polynomial fit, red lines to $4^{th}$ and $6^{th}$ kept constant to the values extracted for SmNiO$_3$, and only the quadratic term fitted to the data.}
    \label{fig:twoPNO}
\end{figure}

The present theory predicts that PrNiO$_3$ is undistorted and metallic, although close to the metal-insulator transition boundary, while the compound is empirically  distorted and insulating, although again with a very low transition temperature. We believe this discrepancy arises because the transition in PrNiO$_3$ is from a paramagnetic metal to an antiferromagnetic insulator, in contrast to the $Sm$ and $Lu$ materials where the transition is between two paramagnetic phases. Inclusion of  antiferromagnetism in the theory will yield a lower energy for the insulating phase of PrNiO$_3$ at the $U$, $J$ values used here while  a slightly different choice of interaction parameters would  also push PrNiO$_3$ to the insulating side of the phase diagram.

We may take the analysis further by substituting the $\Delta N(Q)$ obtained from the solution of Eq.~\ref{EOSnickelate} into the full free energy expression to obtain the free energy as a function of the lattice coordinate $Q$ shown in the right panel of Fig.~\ref{fig:rnoE}. This is analogous to the energy that is usually calculated by electronic structure codes, which work at fixed atomic positions. The discontinuities arise from the different branches of the $\Delta N(Q)$ curves. The resulting free energy is, to a good approximation, the combination of two parabolic energy curves, one corresponding to the metallic state, and one to the insulating state. We see that in this theory the $\Delta N=0$, $Q=0$ state is stable to lattice distortions for PrNiO$_3$, marginally stable for SmNiO$_3$ and unstable for LuNiO$_3$.

%In summary, the data and analysis in this section settles a long-standing debate by showing that the electron-lattice coupling is an essential component of the netal-insulator transition in the rare earth nickelates. While a purely electronically driven transition is in principle possible, it seems not to occur in the bulk perovskite nickelates.

The  analysis presented here is based on a fit of the computed $\Delta N(Q)$ to a free energy model. We emphasize that this fit is in no way essential to our method; one could simply numerically integrate a suitably dense set of $\Delta N(Q)$ data. But for the procedure used here the question of the sensitivity of the results to the choice of free energy arises. To quantify the uncertainties we have refit  the PrNiO$_3$  data shown in Fig. ~\ref{fig:rno1}, now constraining  the $4^{th}$ and $6^{th}$ order coefficients to have the same values as found in SmNiO$_3$. Results are shown in Fig ~\ref{fig:twoPNO}. We see from the left panel that this fit to the data is not quite as good, especially in the small to intermediate $Q$ region. The fitted energy (right panel) is very similar to that shown in Fig. ~\ref{fig:rnoE}, with the value of $\Delta N$ at the minimum almost the same in the two cases and the energy of insulating solution in the constrained fit is now slightly lower than the energy of the metallic solution, placing paramagnetic PrNiO$_3$ slightly on the insulating side of the transition. These differences indicate that the systematic uncertainties in the method are not large.

\begin{figure}[ht]
    \centering
    \includegraphics[width=0.4\textwidth]{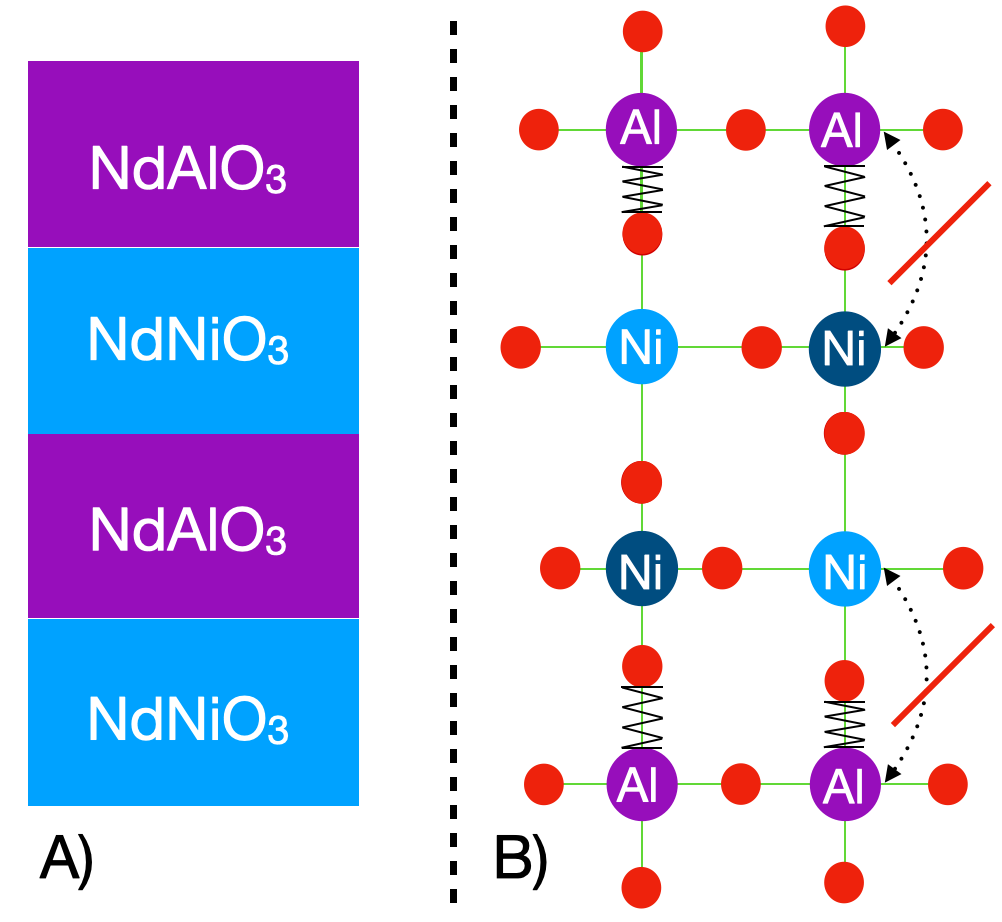}
    \caption{a) Schematic of superlattice consisting of NdNiO$_3$ layered in between the band insulator NdAlO$_3$ b) Schematic view of two atomic layers of NdNiO$_3$ sandwiched between NdAlO$_3$ layers as described in reference \citep{Georgescu2019}. Dashed line with red slash: forbidden hopping path, indicating mechanism of dimensional reduction. Zig-zag lines connecting O and Al: representation of increased stiffness to distortions of the Al-O bonds relative to the Ni-O bonds.}
    \label{fig:nnolayersketch}
\end{figure}

\subsection{Layered Structures of NdNiO$_3$/NdAlO$_3$ \label{subsec:layerednickelates}}

\begin{figure*}
    \centering
    \includegraphics[width=0.9\textwidth]{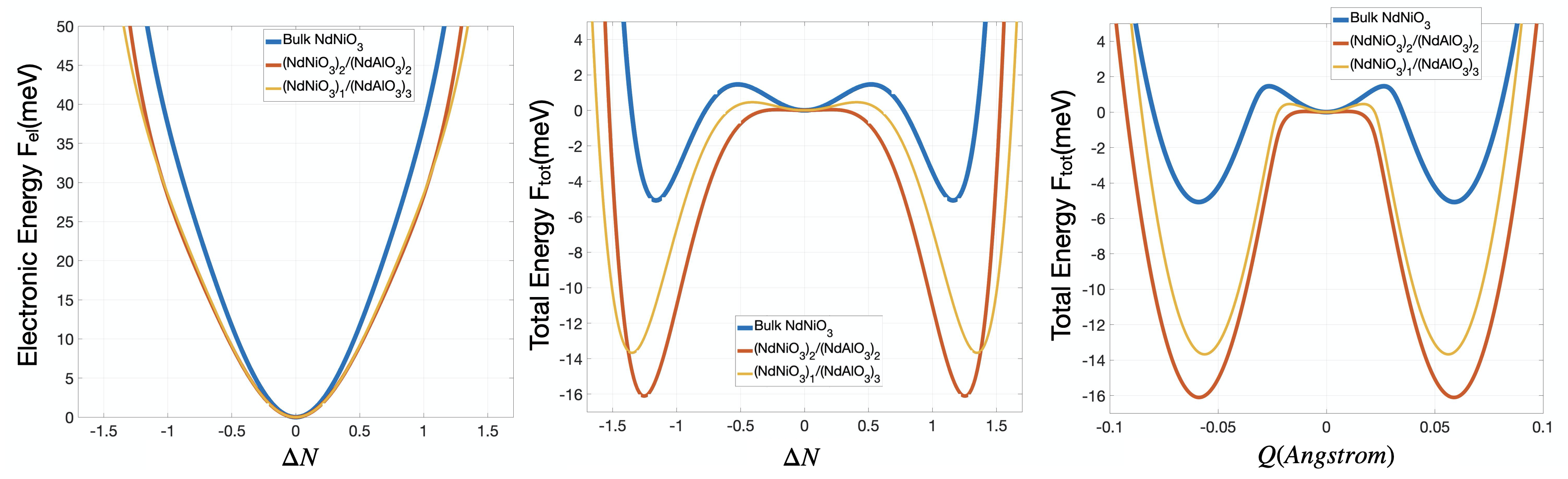}
    \caption{Electronic (left panel)  and total energy  (central panel) as functions of electronic order parameter $\Delta N$ for bulk NdNiO$_3$ and for superlattices consisting of a bilayer of NdNiO$_3$ alternating with NdAlO$_3$ and a monolayer of NdNiO$_3$ alternating with 3 layers of NdAlO$_3$ using data from Ref. \citep{Georgescu2019}. Right panel:  total energy as function of lattice coordinate $Q$ after minimizing over $\Delta N$. Note change of y-axis scale between left-most and other two panels.}
    \label{fig:nnolayer}
\end{figure*}
\begin{figure}[b]
    \centering
    \includegraphics[width=0.4\textwidth]{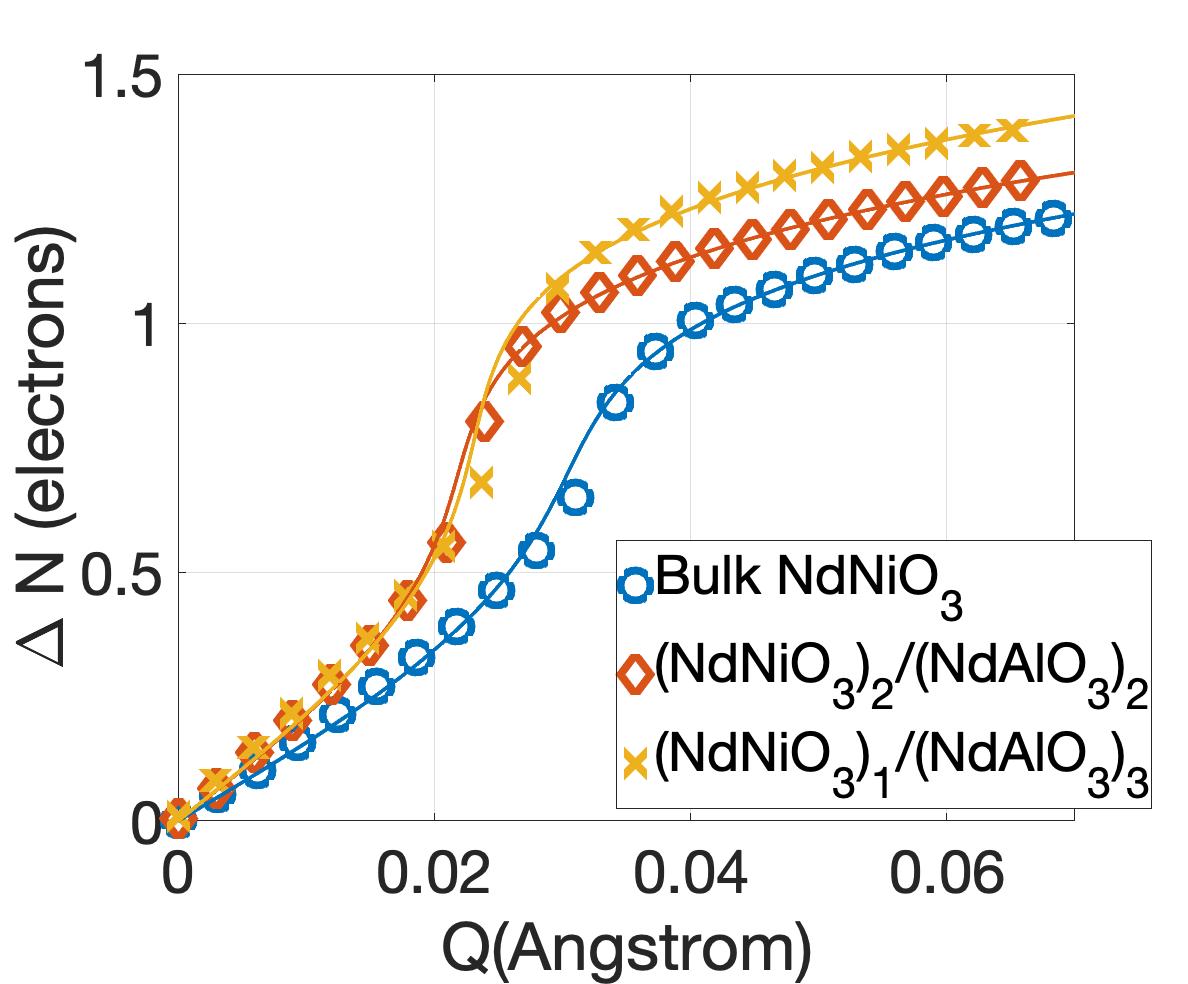}
    \caption{Data from  \citep{Georgescu2019} (points) and the polynomials (light lines) obtained by fitting the data to the equation of state, Eq. \ref{EOSnickelate}}
    \label{fig:Nisuperlattice}
\end{figure}

An important modern direction in quantum materials science is heterostructuring: the ability to synthesize structures in which atomically thin planes of one material alternate with atomically thin planes of another. Previous experimental \citep{Disa2017} and theoretical \citep{Georgescu2019} work analysed systems of the type shown schematically in Fig. ~\ref{fig:nnolayersketch} in which a few layers of metal-insulator transition compound NdNiO$_3$ were grown as a thin film sandwiched epitaxially between layers of the wide-gap insulator NdAlO$_3$.  The few-layer systems had significantly higher transition temperatures than did bulk NdNiO$_3$. Theoretical DFT+DMFT results are available \citep{Georgescu2019} for three systems: bulk NdNiO$_3$ and two heterostructured systems with a 4 unit cell repeat distance: one layer of NdNiO$_3$ followed by three of NdAlO$_3$ (''monolayer"), and two layers of NdNiO$_3$ followed by two of NdAlO$_3$ ("bilayer"). The $\Delta N(Q)$ results are shown in Fig. ~\ref{fig:Nisuperlattice}

Two competing effects were found \citep{Georgescu2019}: the electronic confinement of the electrons in the NdNiO$_3$ by the nearby NdAlO$_3$ layers favors a higher electronic disproportionation $\Delta N$, while the energy cost of imposing a lattice distortion on the adjacent Al-O octahedra is equivalent to an increase in the stiffness $K$ of the NiO$_6$ lattice with a K=31.6914eV/$\AA^2$, for the bulk material, K=35.6269ev/$\AA^2$ for the bilayer and K=40.4753eV/$\AA^2$ for the monolayer; all energies per 10 atom unit cell. The values for g are those quoted in  \citep{Georgescu2019}; we reproduce them here for convenience:  $g=3.2212eV/\AA$ for the bulk, $g=3.3465eV/\AA$ for the bilayer and $g=3.3780 eV/\AA$ for the monolayer.  Performing an analysis similar to that presented in Sec.~\ref{sec:ReNiO3} leads to the energy curves presented in Fig.~\ref{fig:nnolayer} and to the fit parameters shown in Table ~\ref{table:rnolayer}. The original data with the fit are presented for convenience in Fig \ref{fig:Nisuperlattice}

Turning first to the purely electronic correlation energy shown in the left panel of Fig. ~\ref{fig:nnolayer}, we observe that neither NdNiO$_3$ nor the heterostructured materials show a purely electronically driven  insulating state; however, hints of an inflection point, the precursor of the formation of a higher $\Delta N$ extremum, can be seen in bilayer and monolayer curves around $\Delta N=1$ reflecting the effect of quantum confinement in increasing the electron correlations.  Inclusion of the lattice energy produces a global, insulating minimum in all three materials. The monolayer and bilayer show a significantly larger energy difference between the insulating and the metallic states than does bulk NdNiO$_3$, in agreement  with their higher $T_{MIT}$.

\begin{center}
\begin{table}
\begin{tabular}{ | c | c | c |  c | c |}
 \hline
 Layers & $\chi_0^{-1}$(eV) & $\beta $(eV) & $\gamma $(eV) & $-\frac{1}{4}\frac{g^2}{k}$(eV) \\
 \hline
 Bulk & 0.1049 & -0.1022 & 0.0631 & -0.0819 \\ 
 \hline
 Bilayer & 0.0823 & -0.0879 & 0.0545 & -0.0789 \\  
 \hline
 Monolayer & 0.0818 & -0.0747 & 0.0377 & -0.0705  \\  
 \hline
\end{tabular}
\caption{Parameters that characterize the electronic and total energy of NdNiO$_3$ and NdNiO$_3$/NdAlO$_3$ heterostructures consisting of a single or two layers of NdNiO$_3$ from Ref \citep{Georgescu2019}.}
    \label{table:rnolayer}
    \end{table}
\end{center}

The origin of the effect is surprising: the modest decrease in the magnitude of $g^2/K$ as we pass from bulk to bilayer to monolayer reflects the increase in lattice stiffness, opposing the order, and nearly counterbalances the modest decrease in $\chi_0^{-1}$ reflecting the increased correlation physics of the confined metallic state. The more important effect, however, is the change in the magnitude of the higher order ($\Delta N^{4},\Delta N^6$) terms shown in Table ~\ref{table:rnolayer}. The change can be seen    in Fig. \ref{fig:Nisuperlattice} from the variation of the critical $Q$ below which the insulating state is not stable.  In the calculations presented in the previous subsection the higher order terms were also found to be different in PrNiO$_3$ (proximal to the metal-insulator transition) than in LuNiO$_3$ and SmNiO$_3$ (farther from the transition point).

\section{C\MakeLowercase{a}$_2$R\MakeLowercase{u}O$_4$ \label{sec:CaRuO}}

\subsection{Materials and Notation}
\begin{figure}[ht]
    \centering
    \includegraphics[width=0.4\textwidth]{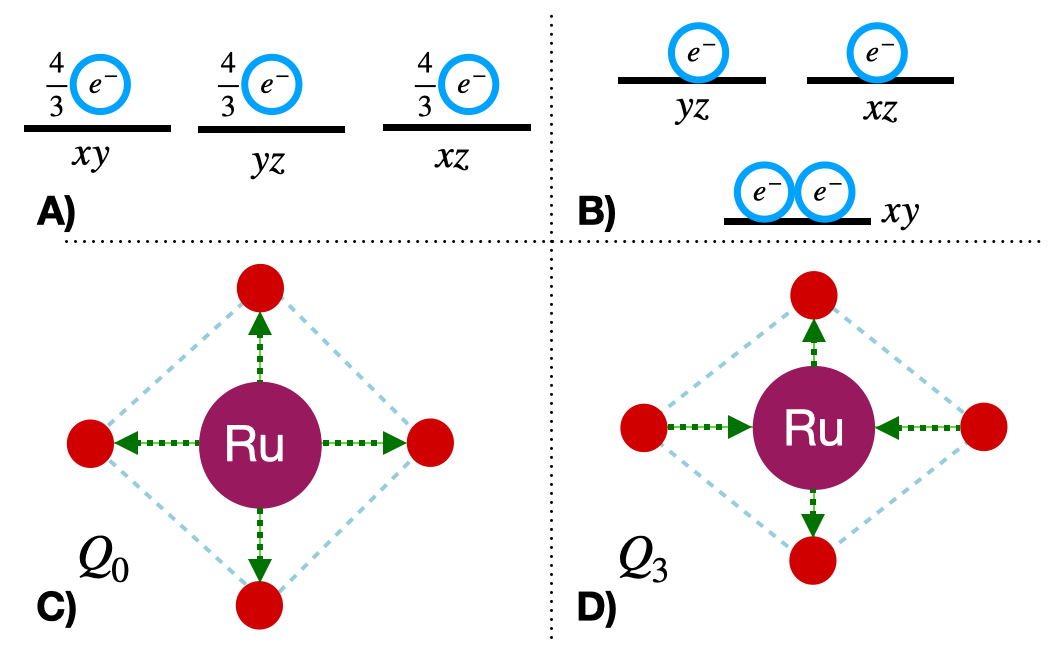}
    \caption{Schematic of the electronic and structural disproportionation characterizing the metal-insulator transition of Ca$_2$RuO$_4$. A) In the metallic phase, all 3 t$_{2g}$ orbitals of Ca$_2$RuO$_4$ are approximately equally occupied; $\Delta N\approx  0$. B) In the insulating phase, the xy orbital is doubly occupied, while the yz and xz are singly occupied; $\Delta N=1$ C) Representation of  Q$_0$ structural mode, in which all bond lengths change equally D) Representation of structural mode $Q_3$ in which octahedral volume is preserved and the change in the two in-plane  Ru-O bond lengths are equal, and opposite in sign to the change in the apical bond length. The electronic state in B) is associated with a structural term of the form $-(Q_3-\lambda_0Q_0)$, with $\lambda_0>0$}
    \label{fig:CROmodes}
\end{figure}

Ca$_2$RuO$_4$ forms in a slightly distorted version of the $n=1$ Ruddlesden-Popper structure, a layered structure consisting of RuO$_2$ planes separated by pairs of CaO layers. Each Ru is six-fold coordinated by oxygen, forming RuO$_6$ octahedra. The relevant electronic states (actually Ru-O hybrids) may be thought of as Ru $t_{2g}$ symmetry d-states ($d_{xy},d_{xz},d_{yz}$) with $4$ $t_{2g}$ electrons per Ru. Below $340K$ the material undergoes a first-order transition from a high temperature metallic to a low temperature insulating phase.  No symmetry is broken at the transition: the two phases share the same crystal structure but differ in the occupancies of the d orbitals and shape of the RuO$_6$ octahedra, and the relative occupancies of the d-levels. As sketched in Fig. ~\ref{fig:CROmodes}, the metallic state is characterized by an approximately equal occupancy of the $t_{2g}$ orbitals while in the insulating state the $xy$ orbitals are doubly occupied and the $xz/yz$ orbitals each contain a single electron. The appropriate electronic order parameter is the occupancy difference between the $xy$ and $xz/yz$ orbitals
\begin{equation}
\Delta N=n_{xy}-\frac{1}{2}\left(n_{xz}+n_{yz}\right).
\label{DeltaNCRO}
\end{equation}
No crystal symmetry protects the orbital occupancies in the high temperature metallic state, which is characterized by a $\Delta N=N_H$ close to, but not exactly, zero.

The change in electron occupancy across the metal-insulator transition occurs simultaneously with a decrease in the Ru-O apical bond length and an increase in the in-plane Ru-O bond length (a flattening of the octahedra). At the transition, there is a change in the lattice constants, and rearrangements of other internal coordinates. The harmonic nature of the lattice Hamiltonian means that most of the lattice modes may be integrated out, leaving an effective theory involving two structural degrees of freedom. The two lattice degrees of freedom are needed, because if the point symmetry of the Ru ion is lower than cubic, then both the unit cell volume change and the relative octahedral distortion couple linearly to the differential level occupancy $\Delta N$.  

Following Ref.~\citep{Han2018} we write the two relevant  structural degrees of freedom as a variable $Q_3$ parametrizing a volume preserving change in the c-direction Ru-O bonds relative to the average in-plane Ru-O bonds, and a change  $Q_0$ in the octahedral volume and define the lattice variables such that in the high-T metallic state $Q_3=Q_0=0$. In terms of changes $\delta x,\delta y, \delta z$ to the three Ru-O bond lengths, defined as:
\begin{equation}
Q_0=\frac{1}{\sqrt{3}}(\delta x+\delta y + \delta z) \: ~~ Q_3=\frac{1}{\sqrt{6}}(2\delta z -\delta x - \delta x)    
\label{eq:QCRO}
\end{equation}

We now build the free energy. We define the lattice distortion relative to the high-T metallic state as the vector $\vec{Q}=(Q_3,Q_0)$; the lattice restoring term $\mathbf{K}$ is a $2\times 2$ matrix with entries determined in previous work\citep{Han2018} to be $K_{33}$=17.7, $K_{03}$=7.6, $K_{00}$=46.2 ev/$\AA^2$ per formula unit. We write the  linear combination of the $\mathbf{Q}$ that couples to the electronic disproportionation as $\vec{\mathcal{F}}\cdot\vec{Q}$ with $\vec{\mathcal{F}}=F_3(1,-\lambda)$. Ref.~\citep{Han2018}  finds F$_3=2.8eV/\AA$ and $\lambda=0.45$. %The metal-insulator order parameter $\Delta N$ was determined to couple to the particular linear combination $Q_3-\lambda_0Q_0$ with $\lambda_0=$ 0.45.  Because no symmetry is broken at the transition the purely electronic free energy may contain odd powers of $\Delta N$. We have:
\begin{equation}
F(\Delta N, Q) =\frac{1}{2}\mathbf{Q}^T\mathbf{K}\mathbf{Q}-\vec{\mathcal{F}}\cdot\vec{Q}(\Delta N-\Delta N_H)+F_{el}(\Delta N),  
\label{FCRO}
\end{equation}
Here $\Delta N_H$ is the value of $\Delta N$ in the high temperature insulating phase; it is nearly, but not quite, zero. 

For the purely electronic energy we chose the $4^{th}$ order form:

\begin{equation}
    F_{el}=\frac{c\Delta N^4}{4}+\frac{b\Delta N^3}{3}+\frac{a\Delta N^2}{2}+\sigma_0\Delta N  
\label{eq:FelCRO}
\end{equation}

Odd powers occur because no symmetry is broken at the transition. 

\subsection{Metal-insulator transition in bulk CaRuO$_4$}

\begin{figure}[ht]
    \centering
    \includegraphics[width=0.4\textwidth]{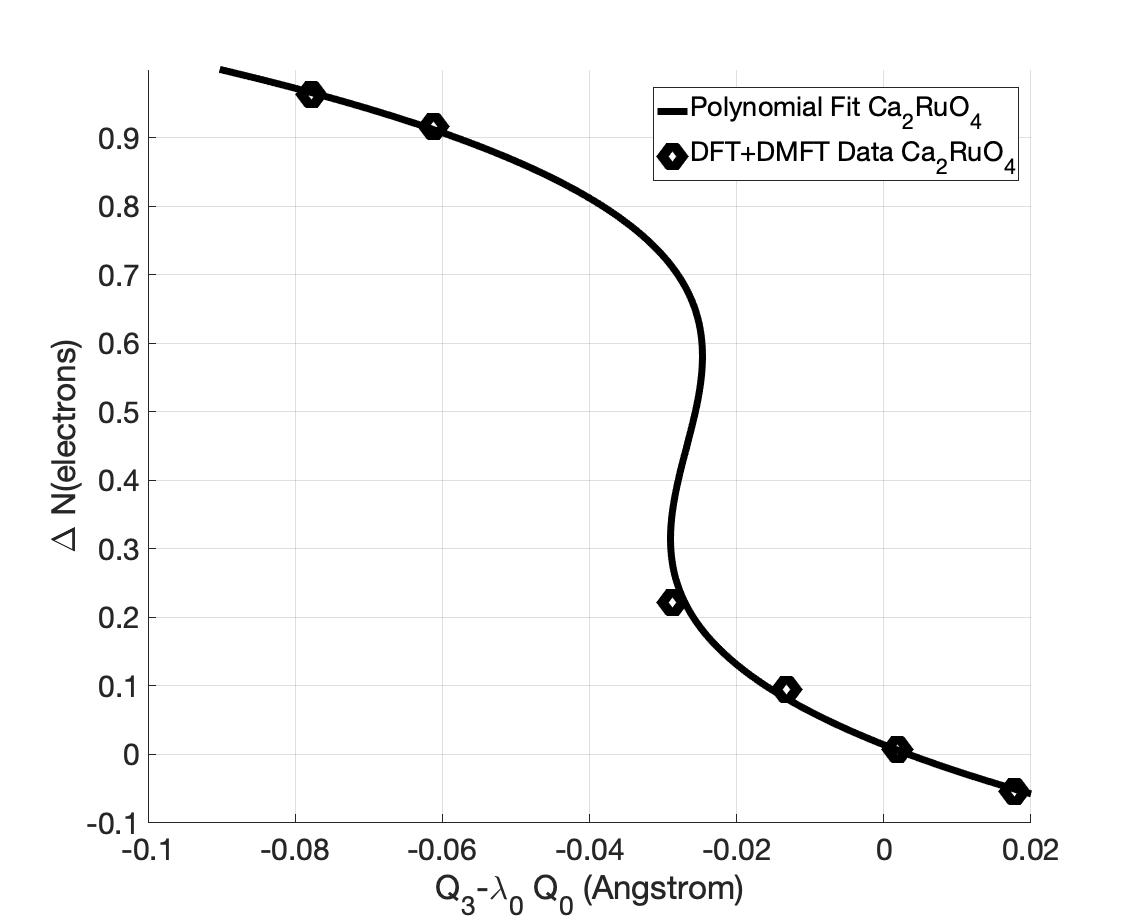}
    \caption{ $\Delta N(Q)$ from \citep{Han2018} (heavy black points) and polynomial fit (solid line) used for the equation of state.}
    \label{fig:CRO}
\end{figure}

Fig. ~\ref{fig:CRO} shows previously published results \citep{Han2018} for $\Delta N$ as a function of the relevant combination of lattice distortions along with our fit to Eq.~\ref{eq:FelCRO}.  As in the previous section we have not used points in the coexistence region for the fit. We find $\sigma_0 = -0.009$, $a = 0.70$, $b = -1.718$, $c=1.28$.

\begin{figure}[ht]  \centering   \includegraphics[width=0.4\textwidth]{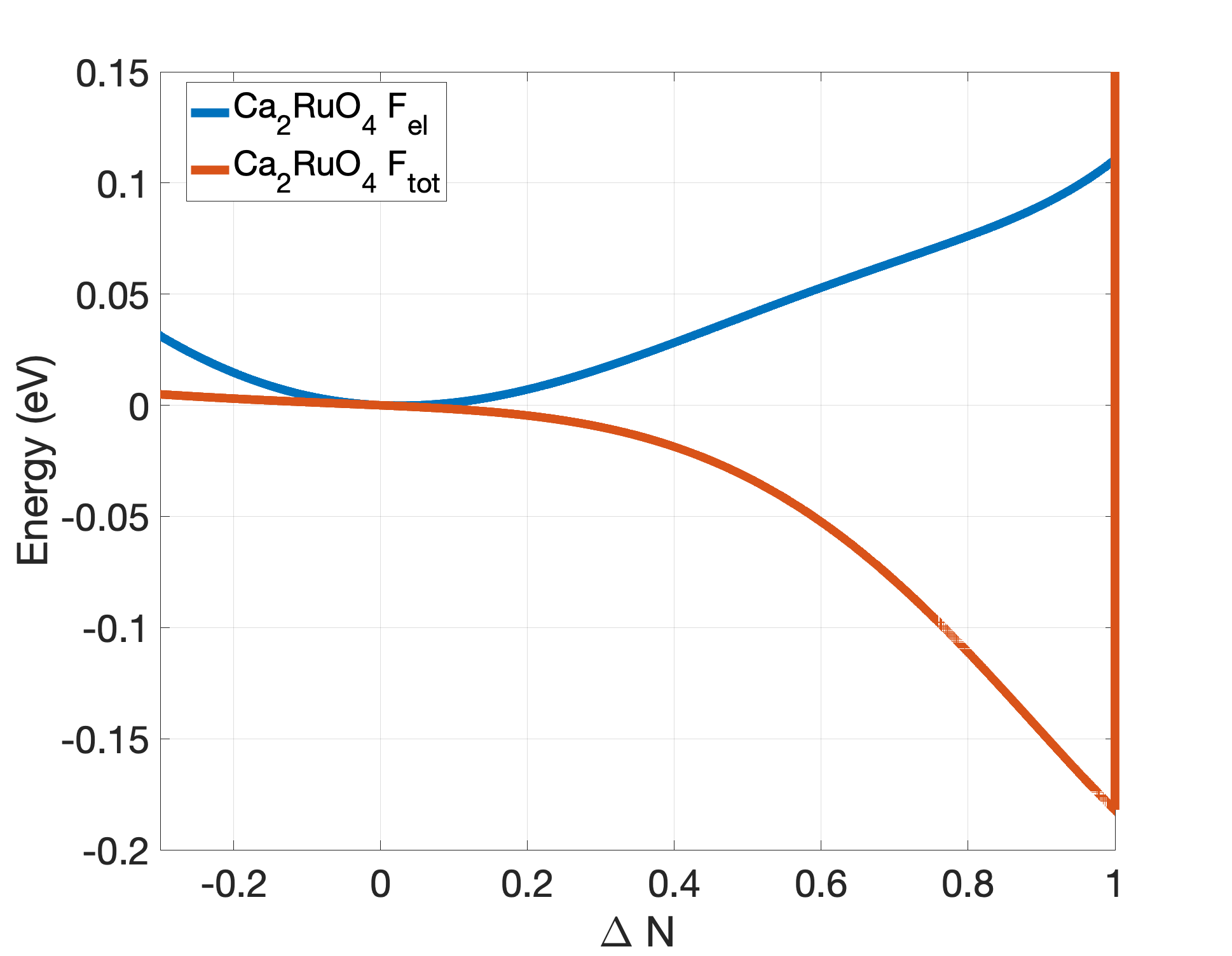}  \caption{$F_{el}(\Delta N)$ and $F_{total}(\Delta N)$  for Ca$_2$RuO$_4$.}   \label{fig:FelCRO}
\end{figure}

The higher curve (blue) in Fig.~\ref{fig:FelCRO} shows the purely electronic free energy resulting from the fit. We see that in the absence of lattice effects the metallic, undistorted state is energetically favored and that there is not even a metastable minimum corresponding to the insulating state although there is a hint of an inflection point that is a precursor of a higher $\Delta N$ extremum. The lower trace (heavy black on-line) shows $\bar{F}$, the full free energy after the lattice modes  has been integrated out. We see that inclusion of  the lattice energetics qualitatively changes the free energy, strongly favoring large $\Delta N$. Indeed we see that the energy does not have a minimum in the physically allowed  range $\Delta N\leq 1$; rather it is minimized at the boundary  $\Delta N= 1$. We again conclude that the transition in this compound is driven by the lattice contribution to the energetics.

%The DMFT computations of  Ref.~\citep{Han2018}, are in effect a theory of the low temperature paramagnetic insulating state and as noted in Ref~\citep{Han2018} the predicted $Q_{3,0}$ correspond well to the extrapolated $T=0$ values of these parameters.   As the temperature is varied from low T to the metal insulator transition temperature $T_{MIT}\approx 340K$, substantial changes are observed in the lattice parameters including the $Q_3$ and $Q_0$\citep{Friedt:CROPT}. These thermally-driven changes are not adequately captured by the existing DMFT calculations. Within the present theory, since in the low t regime $F_3$ is minimized at $\Delta N=1$, for a range of $F_3$ the only source of temperature dependence of the lattice parameters is via a temperature dependence of $F_3$. A similar temperature dependence was  proposed in Ref. \citep{Han2018}. Fig ~\ref{fig:CROMIT} presents calculations of the energy landscape for different values of $F_3$; we see that a first order transition occurs for a $F_3\approx -1.53$  (see Fig.~\ref{fig:CROMIT}  )

\subsection{Epitaxially constrained Ca$_2$RuO$_4$ }

Ca$_2$RuO$_4$ has also been grown expitaxially on insulating substrates. The epitaxial constraint means that the in-plane lattice parameters of the Ca$_2$RuO$_4$ are forced to be equal to the in-plane lattice parameters of the substrate material. {\it A priori}, this constraint does not fix the values of internal coordinates including the  the  Ru-O in-plane bond length of relevance here.  Density functional theory calculations, however, \citep{Han2018} indicate that in practice the material adapts to the epitaxial constraint by adjusting the in-plane Ru-O bond lengths, rather than by rotating the Ru-O$_6$ octahedra, so that the percentage change in the in-plane $Ru-O$ bond lengths is fixed by the percentage change in the in-plane lattice parameters.  The z-direction lattice constants and Ru-O apical bond lengths are still of course free to relax. This means that the theory of the previous section can be simply adapted to the epitaxial case by a change of variables. We write:
\begin{equation}
    Q_3=\frac{1}{\sqrt{6}}(2z-x_0-y_0) 
\end{equation}
and
\begin{equation}
    Q_0=\frac{1}{3}(z+x_0+y_0)
\end{equation}
with $x_0$ and $y_0$ fixed by the epitaxial constraint. It is also convenient to define the zero of $z$ to be the value 
\begin{equation}
z_0=-(x_0+y_0)\frac{K_{00}-K_{33}+\frac{K_{30}}{\sqrt{2}}}{2K_{33}+2\sqrt{2}K_{30}+K_{00}}
\label{z0}
\end{equation}
that $z$ would take at $\Delta N=N_H$.

We obtain:
\begin{equation}
    F(\Delta N,z)=\frac{K^\prime}{2}z^2+g^\prime \left(z-z^\star\right)\Delta N+F_{el}(\Delta N)
    \label{eq:FCRPepi}
    \end{equation}
with
\begin{eqnarray}
K^\prime&=&\frac{2}{3}(K_{33}+\sqrt{2}K_{30}+\frac{1}{2}K_{00})
\\
g^\prime&=&\frac{F_3(2-\sqrt{2}\lambda_0)}{\sqrt{6}}
\label{eq:gprime}
\end{eqnarray}
and
\begin{equation}
z^\star= (x_0+y_0)\left(\frac{1+\sqrt{2}\lambda_0}{2-\sqrt{2}\lambda_0}+\frac{K_{00}-K_{33}+\frac{K_{30}}{\sqrt{2}}}{2K_{33}+2\sqrt{2}K_{30}+K_{00}}\right)
\end{equation}

\begin{figure}[t]
    \centering
    \includegraphics[width=0.5\textwidth]{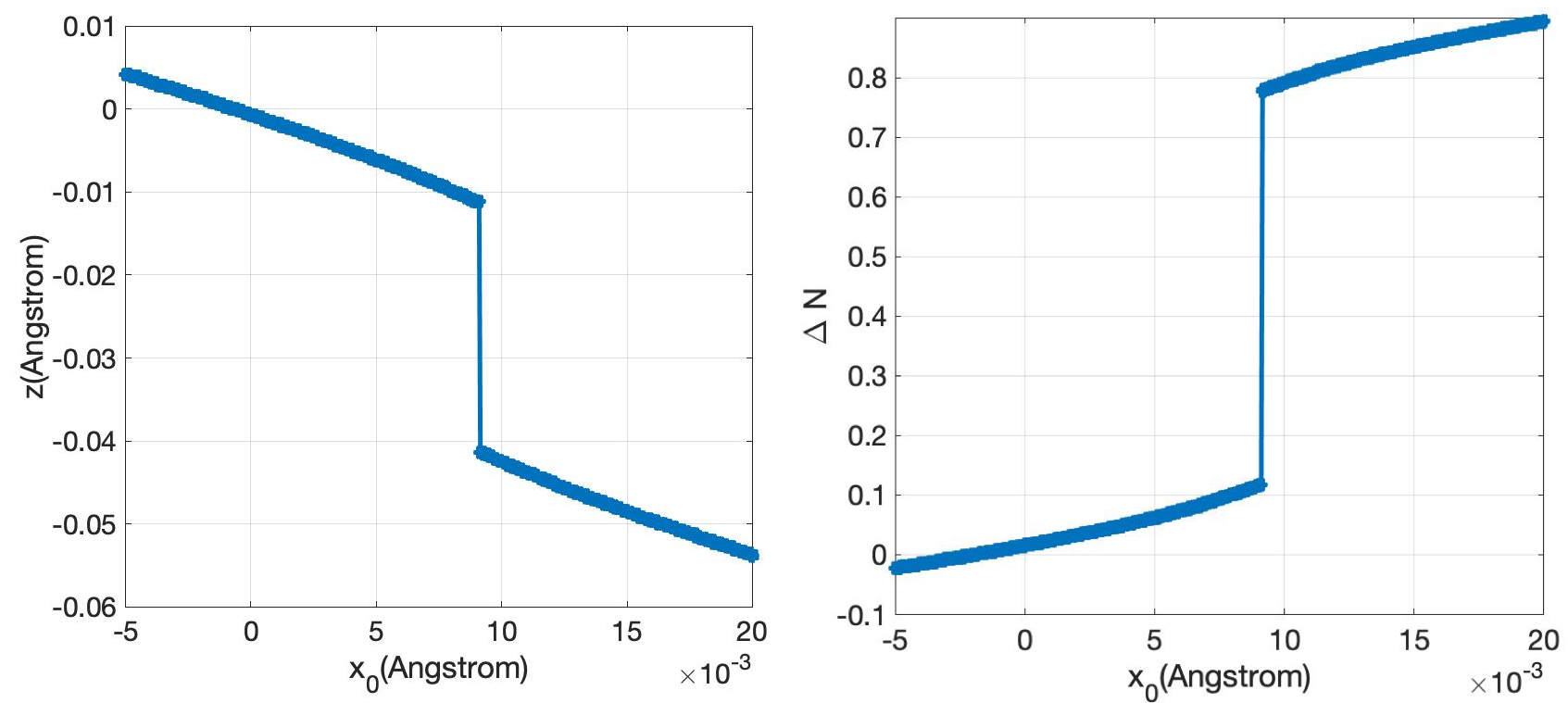}
    \caption{Equilibrium $\Delta N$ and z versus epitaxial strain defined as the difference in mean in-plane Ru-O bond length  relative to the value in the high temperature structure of Ca$_2$RuO$_4$ as imposed through x$_0$+y$_0$ where x$_0$=y$_0$ }
    \label{fig:zndQ}
\end{figure}

\begin{figure}[t]
    \centering
    \includegraphics[width=0.4\textwidth]{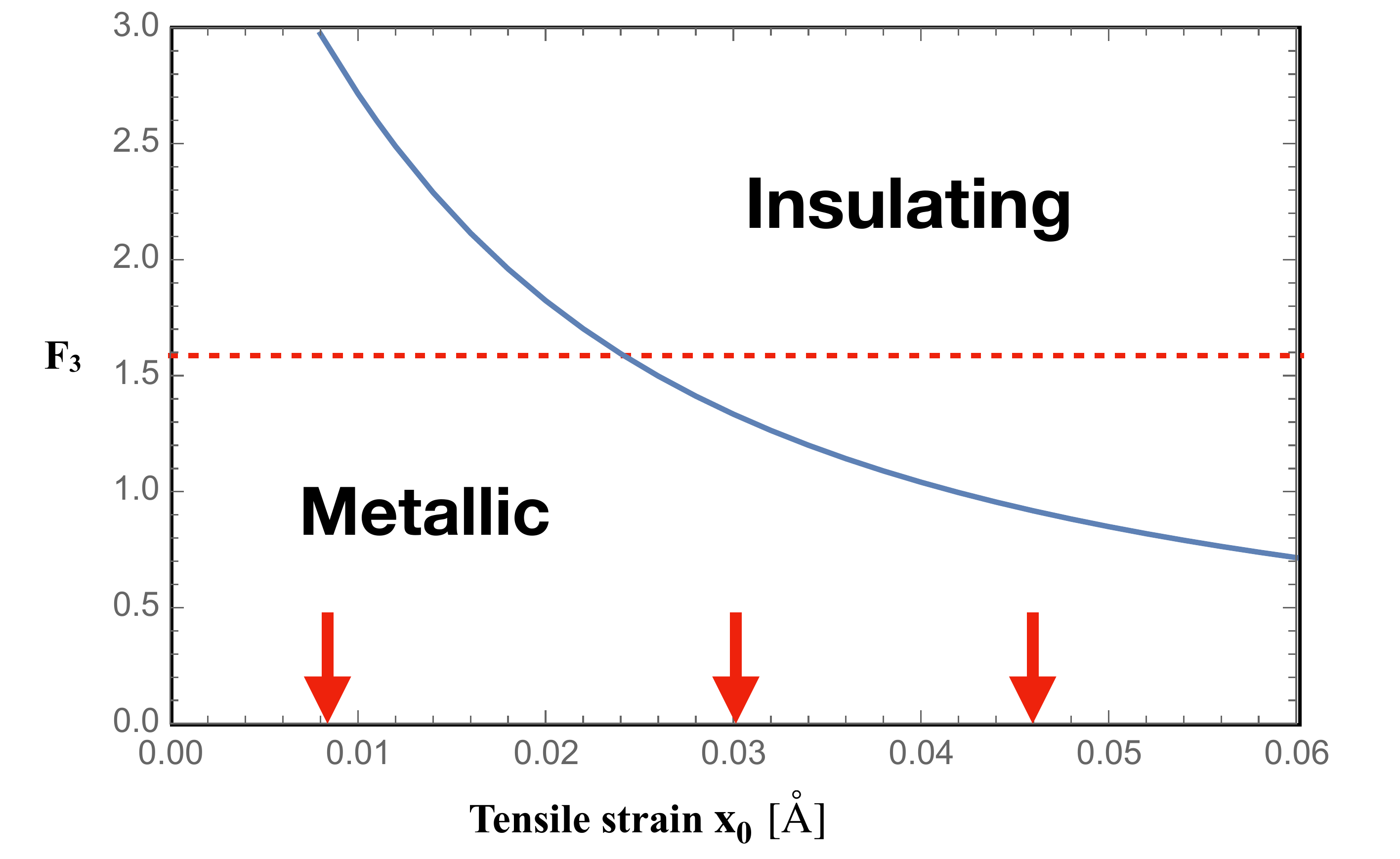}
    \caption{Phase diagram of epitaxially constrained Ca$_2$RuO$_4$ in plane of tensile strain (defined as difference of mean in-plane Ru-O bond length from value in the  high temperature  structure) in \AA \xspace  and coupling parameter $F_3$. Red dashed line: value of $F_3$ at which metal insulator transition occurs in bulk system.  Arrows (red on line)  indicate strain imposed by epitaxial growth on (from left to right) LaAlO$_3$, NSAT and NdGaO$_3$; NdAlO$_3$ corresponding to tensile strain is not shown.}
    \label{fig:CROepiphasediagram}
\end{figure}

\begin{figure}[b]
    \centering
    \includegraphics[width=0.4\textwidth]{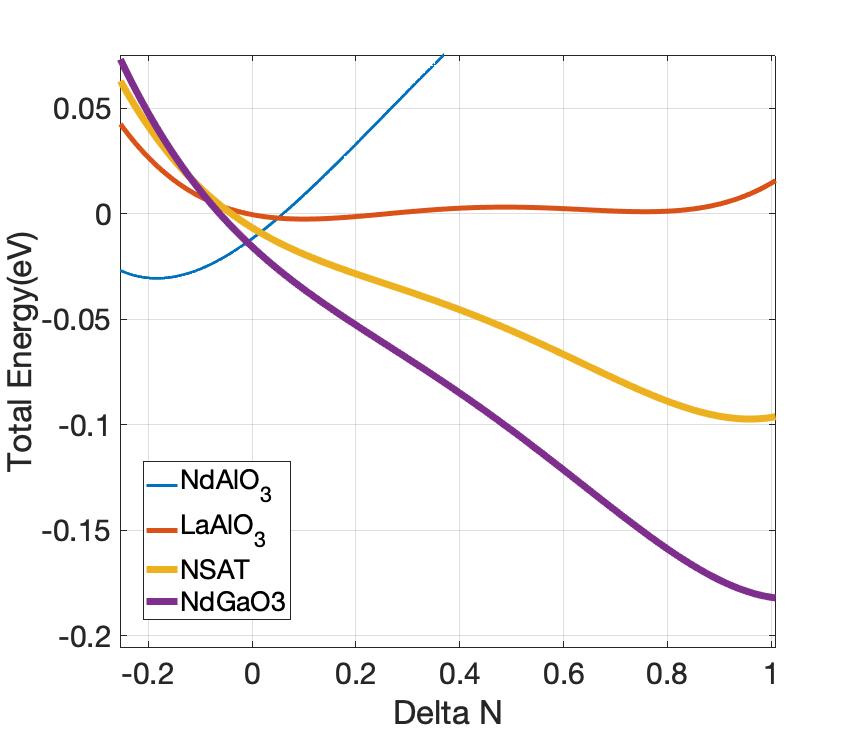}
    \caption{Total energy plotted against orbital disproportionation  $\Delta N$ for materials grown on four substrates indicated in the legend which provide  different strains relative to the metallic phase of Ca$_2$RuO$_4$. For tensile strain (NdAlO$_3$), the material is always metallic and the orbital polarization has the opposite sign from in the insulating state.}
    \label{fig:crostrainlandscape}
\end{figure}
The epitaxial constraint thus has  three effects: it increases the net stiffness to lattice distortions by preventing relaxations of the lattice coordinates associated with in-plane bond lengths, making the insulating phase more difficult to obtain. Second, it reduces the electron lattice coupling (Eq.~\ref{eq:gprime}) which also makes it more difficult to obtain an insulating phase. Finally, it provides a term linearly proportional to $\Delta N$, which for positive epitaxial strain  favors the insulating phase.  The left panel of Fig. ~\ref{fig:zndQ} shows the change in apical Ru-O bond length as a function of epitaxial strain; the right hand panel the occupancy difference $\Delta N$. The first order transition is evident. Note that  for an epitaxial strain matching the in-plane lattice constant of the high temperature metallic phase, insufficient lattice energy is available to stabilize  the insulating phase.  As the tensile strain is increased, the octahedral deformation, which couples linearly to $\Delta N$, increases, and beyond a critical value drives a first order transition. %For epitaxial constraints  such that  the insulating state becomes stable at $T=0$ ($F_3=0$) an insulator-metal transition will occur as temperature is raised. \ajm{As  noted above, decreasing the magnitude of $F_3$ is} a proxy for raising temperature). Decreasing $F_3$ we see that a first order transition to the metallic phase occurs. Fig ~\ref{fig:CROepiphasediagram} shows the phase diagram as a function of in-plane tensile strain and $F_3$. 

Experiments (Ref \citep{CROconference, CROStrain}) have studied films gown epitaxially on substrates NdAlO$_3$, LaAlO$_3$, NSAT and NdGaO$_3$ corresponding   to changes (relative to the high-T state, and assuming the Ru-O bond length exactly tracks the in-plane strain) in average in -plane Ru-O lattice parameters  of $-0.04\AA$, $0.008\AA$, $.03\AA$ and $0.046\AA$ respectively. Fig. ~\ref{fig:CROepiphasediagram} shows that the theory, with no adjustable parameters, predicts that for the theoretically predicted coupling $F_3=2.8eV/\AA$ the films grown on NdAlO$_3$ and LaAlO$_3$ should  be metallic, while the films grown on NSAT and NdGaO$_3$ should be insulating. In the experiment the LaAlO$_3$-strained material has a  transition at $T\approx 200K$ from a high temperature metal to a low temperature weak insulator/bad metal, while the others are metallic and insulating as predicted by the theory. The qualitative agreement is good; the  quantitative discrepancies  may arise from a more subtle relation between Ru-O bond length and in plane lattice constant than was assumed in Ref~\citep{Han2018} or from small systematic errors in the theory. The good correspondence between experiment and theory show the utility of the methodology introduced here. 

\section{Temperature Dependence \label{sec:Tdep}}

The methods presented here provide some insight into the physics of the temperature-driven metal-insulator transition, indicating an important area for future research.   We see from Figs. ~\ref{fig:rnoE} and ~\ref{fig:nnolayer} that for the nickelates the low T energy landscape is characterized by two minima, with energy differences ranging from $80$ meV (LuNiO$_3$) to $25$ meV (NdNiO$_3$).  In the nickelates, as the temperature increases a first order metal-insulator transition occurs. The   lattice distortion $Q$ exhibits negligible temperature dependence over the entire insulating phase, indicating that the minima remain robust and the transition is driven by an entropic effect that lifts up the free energy of the large $\Delta N$ extremum relative to that of the $\Delta N=0$ extremum.

\begin{figure}[ht]
    \centering
    \includegraphics[width=0.5\textwidth]{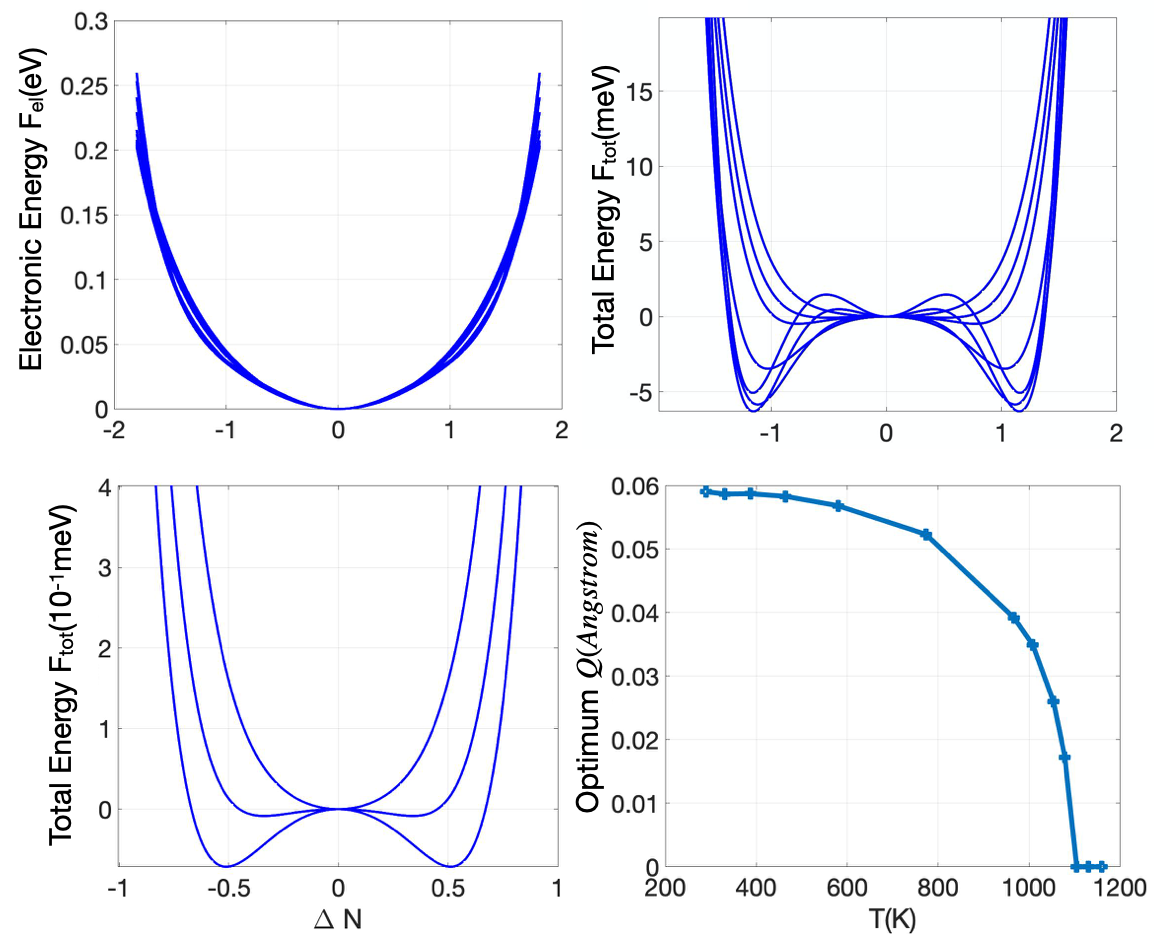}
    \caption{Upper left panel: electronic free energy as function of electronic disproportionation $\Delta N$ computed for NdNiO$_3$ at temperatures from 290K to 1200K. Upper right panel: total free energy computed for bulk NdNiO$_3$ for temperatures of T=290K, 387K, 580K, 773K, 967K, 1054K, 1131K. Lower left panel: total expanded view of total free energy  for bulk NdNiO$_3$ at temperatures of 1054K, 1080K, 1105K. Lower Right Panel: Optimal Q(T)=$\Delta N(T)\frac{g}{2K}$ %with $\Delta N(T)$ the $\Delta N$ value that minimizes  $E_{tot}$ at each $T$.   
    }
    \label{fig:TdepNdNiO3}
\end{figure}

\begin{center}
\begin{table}
\begin{tabular}{ | c | c | c |  c | c | c | }
 \hline
 Temperature  & $\chi_0^{-1}$(eV) & $\beta $(eV) & $\gamma $(eV)  \\
 \hline
 387K &   0.0937 & -0.0781 &  0.0519  \\   
  \hline
580K &  0.0811 & -0.0427  &    0.0346     \\
 \hline
773K &  0.0773 & -0.0197  & 0.0227         \\  
 \hline
966K & 0.0806 & -0.0082 &  0.0172         \\  
 \hline
\end{tabular}
\caption{Parameters that characterize the temperature dependence of the electronic energy  as extracted by fitting the temperature dependent electronic free energy data presented in Fig.~\ref{fig:TdepNdNiO3}}
    \label{table:rnoT}
    \end{table}
\end{center}

To investigate whether the entropic effect arises from the local correlations captured by the dynamical mean field approximation, for bulk NdNiO$_3$ we have extended the calculations shown in Fig. ~\ref{fig:nnolayer} to much higher temperatures by performing calculations with the same methodology and interaction parameters as in \citep{Georgescu2019} but changing the electronic temperature.  Fig.~\ref{fig:TdepNdNiO3}  shows the results: all of the parameters in the electronic free energy vary systematically with temperature, decreasing in magnitude as the temperature is increased from $\sim 290$K to $1000$K; the result is a very weakly first order transition at $\approx 1080$K (the model actually passes very close to the tricritical point at which the second and fourth order coefficients of the theory simultaneously change sign). This behavior, while theoretically interesting, is inconsistent with the observed strongly first order and much lower transition temperature  behavior, suggesting that either the single site dynamical mean field approximation does not adequately describe the temperature dependence of the electronic contribution to the free energy,  or that an entropic effect associated with the lattice degrees of freedom plays an important role. Indeed the neglect of spatial fluctuations generically means that single-site DMFT approximations overestimate transition temperatures \citep{DMFTorderT,DMFTOrderT2}. It should also be noted that in this material antiferromagnetism plays an important role in the ordering, and that a direct computation \citep{Haule2017Nickelates} of the free energy $F(\Delta N,Q)$  using a different DMFT formalism places paramagnetic  NdNiO$_3$ on the paramagnetic metal side of the transition and yields a stronger low T temperature dependence, and shows for LuNiO$_3$ a temperature dependence of $F(\Delta N,Q)$ in the range $T<500K$ very similar to that found here for NdNiO$_3$.

\begin{figure}[t]
    \centering
    \includegraphics[width=0.4\textwidth]{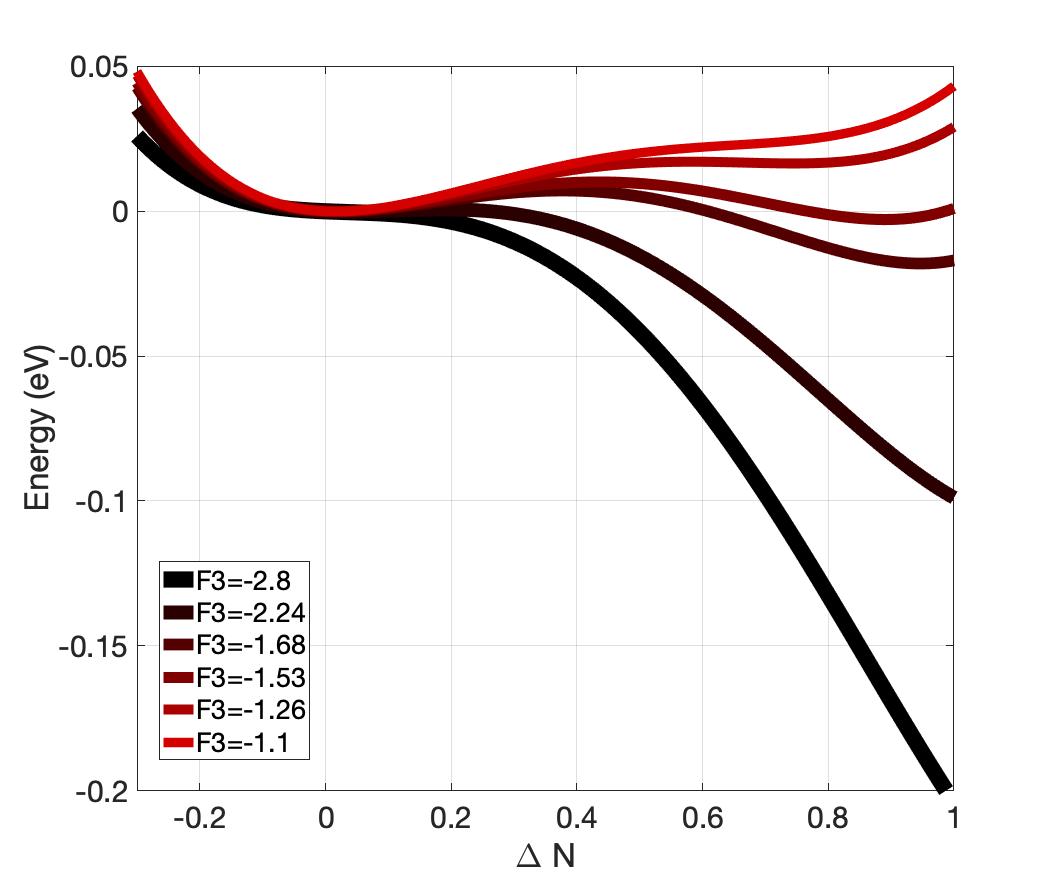}
    \caption{$F_{total}(\Delta N)$  computed for bulk Ca$_2$RuO$_4$ for different values of F$_3$. }
    \label{fig:CROMIT}
\end{figure}

Ca$_2$RuO$_4$ presents different issues. A substantial temperature dependence is observed \citep{Friedt:CROPT} across the insulating phase. Since the electronic order parameter $\Delta N$ in this material is fixed at its largest possible value for a range of couplings, the temperature dependence must imply a temperature dependence of the electron-lattice coupling $F_3$ as proposed for related reasons in \citep{Han2018}. Within single-site dynamical mean field theory, there is no theoretical justification for a strongly temperature-dependent coupling parameter, but on the assumption that changes in electron-lattice coupling F$_3$ are a proxy for changes in temperature we show in Fig. ~\ref{fig:CROMIT} the energy computed for different values of $F_3$. We see that a first order transition occurs at an $F_3\approx 1.58\approx 0.6 $ of the $F_3$ we estimated from the low temperature calculation. The phase diagram presented in Fig.~\ref{fig:CROepiphasediagram} then indicates that films grown on substrates with a tensile epitaxial constraint exhibit significantly higher transition temperatures than found in bulk, roughly consistent with experiment.

\section{Discussion and Outlook \label{sec:summary}}

This paper has analysed metal-insulator transitions in quantum materials by constructing an energy functional $F(\Delta N,Q)$ that depends on suitably defined electronic ($\Delta N$)   and lattice ($Q$) order parameters. The first  assumption enabling the construction is that the lattice energy is quadratic in deviations from an equilibrium position at fixed value of the electron order parameter, in other words that the physics controlling the relevant atomic force constants arises from chemical bonds inside the solid that are only weakly affected by the transition from metal to insulator. This and the harmonic nature of the lattice response are found in density functional calculations (see e.g. the supplementary material of Ref. ~\citep{Georgescu2019}) and are further supported by the observation that phonon frequencies do not change much across the transitions of interest. The second assumption is that the coupling between electronic and lattice degrees of freedom is linear, and may be determined a-priori.   This second assumption  has been confirmed by explicit calculations (see, for example, the supplementary information in  \citep{Georgescu2019}).  Further investigation of these assumptions, and identification of compounds in which they break down, is an important topic for future research. One important direction is a detailed comparison of results obtained along the lines indicated here to direct computations of energies and free energies \citep{Park2013,Park2014,Haule2015,Haule2017Nickelates} 

Given these assumptions, a many-body calculation of electronic configuration as a function of lattice  distortion may be used to construct free energies, essentially by integrating the equation of state. We accomplish the integration by fitting the equation of state results to a polynomial which is integrated analytically. This procedure avoids ambiguities associated with the coexistence region of the first order transition. 

We applied the methodology to two currently interesting families of compounds, that exhibit first order metal-insulator transitions as the temperature is decreased: the rare earth perovskite nickelates RNiO$_3$, and Ca$_2$RuO$_4$.  The transition in the RNiO$_3$ is a symmetry breaking transition leading to two inequivalent Ni sites and an alternating pattern of long and short Ni-O bond lengths. The transition in Ca$_2$RuO$_4$ is isosymmetric, with no broken crystal symmetries. In both materials, we find that if the lattice coordinate is set to the value appropriate to the high temperature metallic state, the resulting energy $F(\Delta N,Q=0)$ has only one minimum, at $\Delta N=0$. It is only after the coupling to the lattice is included that the total functional $F(\Delta N,Q)$ acquires a minimum at a $\Delta N,Q\neq 0$. We therefore conclude that lattice effects are essential in driving the metal-insulator transition in both compounds, rather than being merely a minor consequence of a fundamentally electronically driven transition, thereby settling a long-standing controversy. It is important to note however that although the purely electronic theory does not produce a metal-insulator transition,  the correlation contribution to the electronic energy is still important and calculations on a beyond-DFT level are apparently required to obtain  correct  energetics. 

It is important to consider the limitations of our conclusions. First, while the  approach introduced here will work with any method that calculates the electronic order parameter as a function of lattice distortion $\Delta N(Q)$, in practice the information available in the literature comes from the density functional plus dynamical mean field methodology. While this method  successfully produces results that are consistent with many experiments, its precise quantitative accuracy is unknown. The method does require an  identification of correlated orbitals,  it relies on the assumption that density functional theory gives an adequate account of the energetics of the uncorrelated orbitals, it requires specification of interaction parameters, and  its solution of the many body problem involves a stringent locality assumption.  Cross checking via other methods (as for example was done for Ca$_2$RuO$_4$ in Ref. \citep{CRO2020} ) would be desirable.  Further, there are several variants of the DMFT plus DFT methodology, differing the choice of energy window and the representation of correlated states \citep{karp2021dependence}; the dependence of our conclusions on the implementation is an interesting question.

The specification of interaction parameters is of particular importance. As shown e.g. in Refs.~\citep{Georgescu2020,Subedi2015}, as the interaction parameters are increased in magnitude, a transition can be generated in the purely electronic theory. The precise statement made here is that with interaction parameters determined by reproducing  physical properties including gaps and effective masses , the purely electronic theory does not exhibit a phase transition: electron-lattice coupling plays an essential role in stabilizing the insulating states. 

It will also be noted that while the theory presented here reproduces trends and orders of magnitude very well, it has some quantitative deficiencies, for example predicting that PrNiO$_3$ is metallic but close to the phase boundary when in fact it is insulating but close to the phase boundary, with the lowest transition temperature of any bulk member of the $R$NiO$_3$ nickelate family. This deficiency is likely to be remedied by inclusion of magnetism in the theory, since in PrNiO$_3$ and NdNiO$_3$, but not in the other compounds studied in this paper, the metal-insulator transition is accompanied by a magnetic transition.   Similarly the phase diagram of epitaxially constrained Ca$_2$RuO$_4$ films is qualitatively correct but exhibits  quantitative discrepancies with experiment. We note that in this paper no attempt was made to adjust the parameters from the literature or to fine-tune the fits  to obtain better agreement with experiment. Small changes, reflecting moderate parameter uncertainties and modest systematic errors in the DFT+DMFT approach, would likely cure these discrepancies.

The  access to the free energy provided by our methods enables additional insights. Issues relating to temperature dependence were discussed in Section ~\ref{sec:Tdep}. The energy landscape shown in Fig. ~\ref{fig:rnoE1} makes it clear that only one narrowly defined path connects the metallic and insulating minima; this information may be used to investigate the kinetics of order parameter nucleation if the material is supercooled or superheated across the phase boundary. Further, the differing roles of the electronic and lattice degrees of freedom in defining the energy landscape make possible an analysis of the kinetics of state evolution after excitation. For example, excitation might rapidly heat the electrons (leading to changes in $F_{el}$) but only slowly heat the lattice degrees of freedom (which would in any event respond more slowly).  Extension of the theory presented here to include the  antiferromagnetism that occurs in the rare earth nickelates may help understand recent ultrafast experiments exploring the relation between antiferromagnetism, lattice distortions and metallicity in NdNiO$_3$ \citep{Stoica2020,Forst2015a}.

The methods can be applied to quantify the importance of lattice effects in other systems of current interest, including CaFeO$_3$  \citep{CFOOriginal,CFO} which exhibits charge ordering and lattice disproportionation, perovskite titanates and vanadates, where lattice distortions couple to orbital ordering \citep{Pavarini2004,Beck2020,Zhang2020}, VO$_2$ where a dimerization instability occurs \citep{VO2Biermann},  V$_2$O$_3$ where the metal-insulator transition couples to the volume of the material \citep{Leonov2015}, and the rare earth manganites where electronic charge, orbital and magnetic ordering are tightly coupled to lattice distortions \citep{Manganite1,McLeod2020}. We also observe that while we presented results based on dynamical mean field calculations, the method is generic in nature and can easily be used with other electronic structure methods, including DFT, DFT+U, Gutzwiller\citep{LDAGutzwiller,Gutzwiller}, Auxiliary-Boson\citep{Lau2013,DeMedici2017,Florens2004,Georgescu2015} - particularly with the advent of new codes that calculate energy and naturally handle symmetry breaking\citep{BOSS,Georgescu2017,SlaveVariation,Hassan2010}, DFT+U\citep{Anisimov1991,Anisimov1997}, and quantum Monte Carlo\citep{QMC}.

\begin{acknowledgments}
We are indebted to Antoine Georges and Oleg Peil for inspiration,  helpful comments and advice at various stages of this work.  Discussions with Jennifer Fowlie, Claribel Dominguez, Bernat Mundet Bolos, Marta Gibert, and Jean-Marc Triscone stimulated this project. We would also like to acknowledge suggestions to improve this manuscript from Lauren Walters, Sophie Beck, Alexander Hampel, Claribel Dominguez, as well as conversations with James Rondinelli, Kyle Miller and Danilo Puggioni during the preparation of this manuscript. ABG was partly supported by the Advanced Research Projects Agency-Energy (ARPA-E), U.S.\ Department of Energy, under Award Number DE-AR0001209. The views and opinions of authors expressed herein do not necessarily state or reflect those of the United States Government or any agency thereof. The Flatiron Institute is a division of the Simons Foundation.
\end{acknowledgments}
%\begin{Appendix}
%\section{Appendix}
%Here we show the data used to fit the electronic equation of state for the NNO/NAO superlattices, and the resulting fit. The data is the result of calculations from Ref. \citep{Georgescu2019}

%\end{Appendix}
%\nocitep{*}

\bibliography{Mendeley}% Produces the bibliography via BibTeX.

\end{document}